\documentclass[a4paper,english,reprint, aps]{revtex4-1}
\usepackage[T1]{fontenc}
\usepackage[utf8]{inputenc}
\usepackage{float}
\usepackage{textcomp}
\usepackage{amsmath}
\usepackage{amssymb}
\usepackage{graphicx}

\makeatletter

\pdfpageheight\paperheight
\pdfpagewidth\paperwidth

\usepackage{lmodern}
\usepackage[T1]{fontenc}

\makeatother

\usepackage{babel}
\begin{document}

\title{Self-referencing of an on-chip soliton Kerr frequency comb \\ without
external broadening}

\author{Victor Brasch}

\altaffiliation{These authors contributed equally to this work.}

\affiliation{École Polytechnique Fédérale de Lausanne (EPFL) -- Institute of Physics,
Lausanne, CH-1015, Switzerland.}

\author{Erwan Lucas}

\altaffiliation{These authors contributed equally to this work.}

\affiliation{École Polytechnique Fédérale de Lausanne (EPFL) -- Institute of Physics,
Lausanne, CH-1015, Switzerland.}

\author{John D. Jost}

\affiliation{École Polytechnique Fédérale de Lausanne (EPFL) -- Institute of Physics,
Lausanne, CH-1015, Switzerland.}

\author{Michael Geiselmann}

\affiliation{École Polytechnique Fédérale de Lausanne (EPFL) -- Institute of Physics,
Lausanne, CH-1015, Switzerland.}

\author{Tobias J. Kippenberg}

\affiliation{École Polytechnique Fédérale de Lausanne (EPFL) -- Institute of Physics,
Lausanne, CH-1015, Switzerland.}

\begin{abstract}
Self-referencing turns pulsed laser systems into self-referenced frequency combs.
Such frequency combs allow counting of optical frequencies and have
a wide range of applications. The required optical bandwidth to implement
self-referencing is typically obtained via nonlinear broadening in
optical fibers. Recent advances in the field of Kerr frequency combs
have provided a path towards the development of compact frequency
comb sources that provide broadband frequency combs, exhibit microwave
repetition rates and that are compatible with on-chip photonic integration.
These devices have the potential to significantly expand the use of
frequency combs. Yet to date self-referencing of such Kerr frequency
combs has only been attained by applying conventional, fiber based
broadening techniques. Here we demonstrate external broadening-free
self-referencing of a Kerr frequency comb. An optical spectrum that
spans two-thirds of an octave is directly synthesized from a continuous
wave laser-driven silicon nitride microresonator using temporal dissipative
Kerr soliton formation and soliton Cherenkov radiation. Using this
coherent bandwidth and two continuous wave transfer lasers in a \textit{2f-3f}
self-referencing scheme, we are able to detect the offset frequency
of the soliton Kerr frequency comb. By stabilizing the repetition
rate to a radio frequency reference the self-referenced frequency
comb is used to count and track the continuous wave pump laser's frequency.
This work demonstrates the principal ability of soliton Kerr frequency
combs to provide microwave-to-optical clockworks on a chip.
\end{abstract}


\maketitle

Nonlinear spectral broadening in optical fibers via supercontinuum
generation \citep{Ranka2000,Dudley2006,Skryabin2010} can provide
an octave of coherent optical bandwidth from pulsed lasers \citep{Cundiff2003}.
This discovery has been an essential step in realizing self-referencing
schemes, which enable measuring the carrier envelope offset frequency
of a frequency comb \citep{Jones2000,Reichert1999}. By measuring
the offset frequency ($f_{\text{{CEO}}}$) and the repetition rate
($f_{\text{{rep}}}$), the optical frequencies of all the comb lines
can be precisely determined via the relation $\nu_{n}=f_{\text{{CEO}}}+n\cdot f_{\text{{rep}}}$
where $n$ designates the respective comb line. This relation establishes
a phase coherent link from the radio frequency (RF) to the optical
domain and provides a ``clockwork'' that enables counting of optical
frequencies \citep{Cundiff2003,Udem2002} or, in the reverse direction,
the synthesis of optical frequencies from radio frequencies \citep{Holzwarth2000}.
These properties have made self-referenced frequency combs versatile
precision tools for many applications such as optical atomic clocks
\citep{Diddams2001}, spectroscopy \citep{Newbury2011} and low-noise
microwave generation \citep{Fortier2011}. 

\begin{figure}[h!]
\includegraphics[width=0.95\columnwidth]{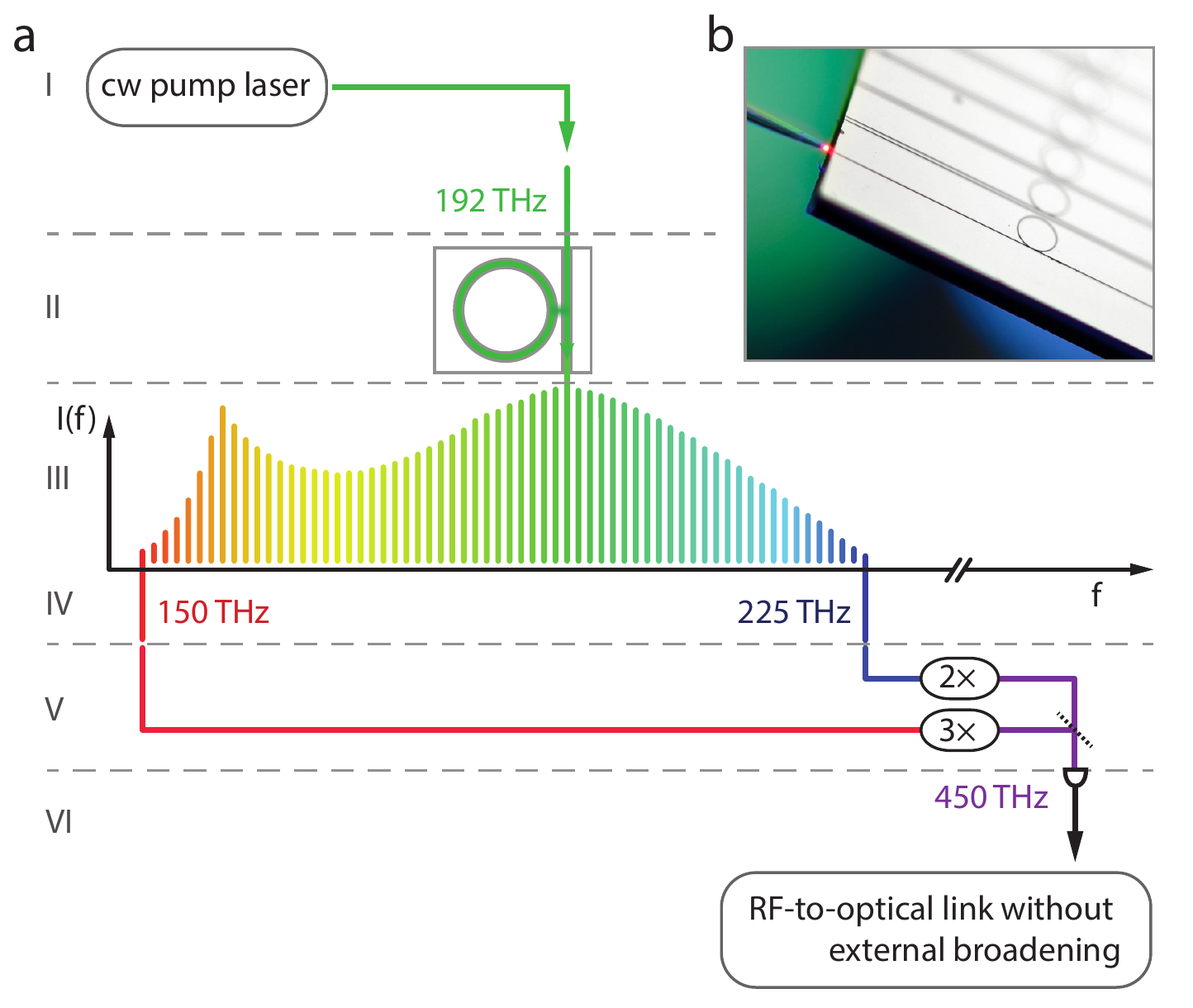}

\caption{\textbf{Schematic illustration of a self-referenced Kerr frequency
comb. a} The soliton Kerr frequency comb is generated from a continuous
wave laser (I) coupled to a silicon nitride microresonator (II). With
a spectral width of two-thirds of an octave the resulting Kerr frequency
comb (III) is sufficiently broad to allow for the measurement of its
offset frequency (i.e. the carrier envelope offset frequency). To
do so, two transfer lasers (IV) are phase locked to the frequency
comb and doubled and tripled in frequency via nonlinear crystals (V).
The heterodyne beat note of the two lasers allows the measurement
of the offset frequency ($f_{\text{{CEO}}}$) of the Kerr frequency
comb (VI). \textbf{b} A photograph of a chip with integrated silicon
nitride microresonators (240 \textmu m diameter) and bus waveguides.
Also shown is a lensed fiber used to couple light (in this photograph
a red laser) into the waveguides on the chip.}
\end{figure}

\begin{figure*}
\includegraphics[width=1.98\columnwidth]{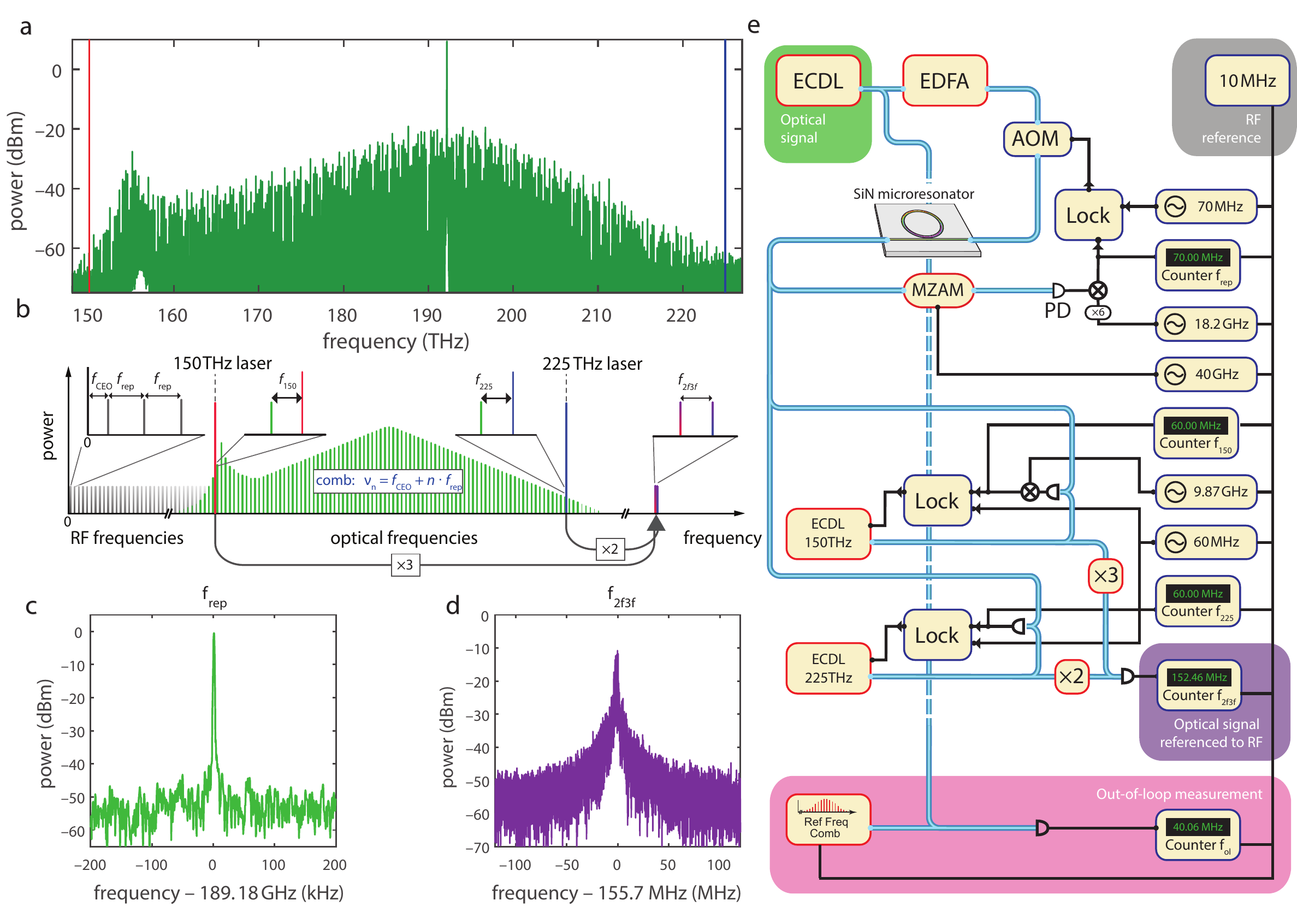}

\caption{\textbf{Setup and employed scheme for self-referencing of a soliton
Kerr frequency comb. a} Optical spectrum of the four-soliton state
that was self-referenced in this work. \textbf{b} Schematic of the
\textit{2f-3f} self-referencing with two transfer lasers as it was
used in this work including the two stabilized frequency offsets $f_{150}$
and $f_{225}$ of the transfer lasers. \textbf{c} In-loop beat note
of the stabilized repetition rate centered at 189.2 GHz. \textbf{d}
Measurement of the free-running $f_{2f3f}$ beat note which allows
the computation of the offset frequency of the Kerr frequency comb.
\textbf{e} Setup used for the experiments. AOM: acousto-optic modulator;
ECDL: external cavity diode laser; EDFA: erbium-doped fiber amplifier;
LOCK: combined phase comparator and proportional-integral-derivative
(PID) servo controller for the phase locks; MZAM: Mach-Zehnder amplitude
modulator; PD: photodiode. All RF frequencies are derived from one
RF reference that is also used to reference the measurements. Blue
lines represent optical fibers, black lines represent electrical connections.}
\end{figure*}

The discovery of Kerr frequency comb generation in optical microresonators
\citep{DelHaye2007,Kippenberg2011} (also known as microresonator
frequency combs), has triggered substantial research efforts towards
the development of compact frequency comb sources with repetition rates in
the microwave regime (>10 GHz) and spectral operation from the near-infrared
\citep{Savchenkov2011} to the mid-infrared \citep{Wang2013,Griffith2015},
which are fully compatible with on-chip photonic integration \citep{Moss2013}.
In the context of frequency metrology one advantage of Kerr frequency
combs is that due to their high repetition rates the line spacing is
usually sufficiently large to be resolvable on lower resolution grating-based
optical spectrometers. This greatly simplifies applications where
the knowledge of the line number $n$ is required \citep{Newbury2011}
or where the frequency comb is used to calibrate a spectrometer \citep{Steinmetz2008}.
Kerr frequency combs have the potential to significantly extend the utility
and range of applications of frequency combs by reducing size, complexity
and costs, and they have already been successfully employed for a
range of applications including coherent terabit communications \citep{Pfeifle2014},
atomic clocks \citep{Papp2014} and optical arbitrary waveform generation
\citep{Ferdous2011}. The recent observation of Kerr frequency combs
generated via dissipative temporal soliton formation \citep{Herr2013,Leo2010}
has been a pivotal development. Dissipative solitons in microresonators
provide a reliable path towards fully coherent comb operation as well
as access to femtosecond optical pulses at microwave repetition rates.
Such dissipative Kerr solitons rely on the double balance of parametric
gain and cavity loss, as well as of Kerr nonlinearity and dispersion
\citep{Leo2010,Matsko2011,Coen2013,Herr2013} and have been generated
in a number of microresonator platforms to date \citep{Herr2013,Liang2015,Brasch2016,Yi2015,Joshi2016}.

Dissipative Kerr solitons in optical microresonators also provide
a route to synthesize spectra that are sufficiently broad for self-referencing
without the need of external broadening similar to Ti:sapphire lasers
\citep{Ell2001,Fortier2003} and in contrast to previous demonstrations
of self-referenced Kerr frequency combs that relied on both external
amplification and broadening stages \citep{Jost2015,DelHaye2015}.
Superseding these stages makes further on-chip integration of self-referenced
Kerr frequency comb sources possible which could enable the realization
of a fully chip-scale RF to optical link. The required optical bandwidth
for self-referencing is achieved due to the large cavity enhancement
along with dispersion engineering \citep{Okawachi2014} in photonic
chip-based silicon nitride (Si$_{3}$N$_{4}$) microresonators. This
allows the generation of solitons for which the Raman effect \citep{Milian2015,Karpov2016,Yi2015}
and higher order dispersion effects such as soliton Cherenkov radiation
\citep{Jang2014,Akhmediev1995} (a process related to third order
dispersion and also known as dispersive wave emission) become relevant.
Recent results have shown that the generation of coherent spectra
spanning two-thirds of an octave is possible using dispersive wave
emission \citep{Brasch2016}. 

\begin{figure}
\includegraphics[width=0.9\columnwidth]{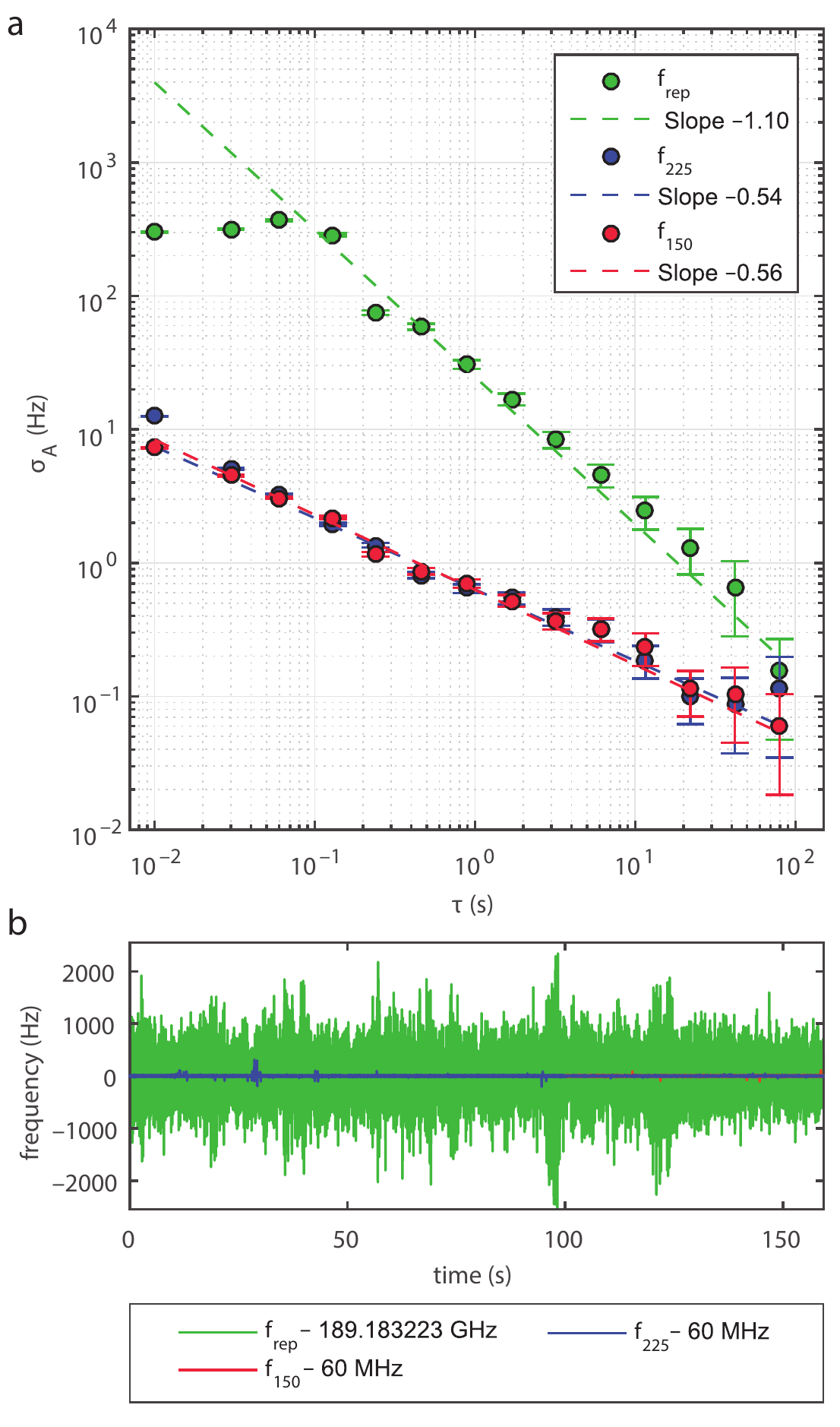}

\caption{\textbf{Counter measurements of the stabilized beat frequencies. a}
The overlapping Allan deviation $\sigma_{\text{{A}}}$ of all locked
frequencies (the repetition rate of the Kerr frequency comb $f_{\text{{rep}}}$
and the offsets of the two transfer lasers $f{}_{150}$ and $f_{225}$)
average down for longer gate times, showing all frequencies are indeed
phase locked. The flat part of the Allan deviation for $f_{\text{{rep}}}$
is caused by the limited bandwidth of the actuation for this phase
lock. The two transfer laser frequency offsets $f_{225}$ and $f_{150}$
are counted with different frequency counters. The beat $f_{150}$ is counted
on a counter with dead time resulting in a slope of only around --0.5
instead of --1. The beat $f_{225}$ is counted on a gapless counter.
The reason for the deviation from the ideal $\tau^{-1}$ behavior
for $f_{225}$ are the occasional frequency excursions that the lock
does not compensate perfectly. \textbf{b} The time trace of the locked
and counted signals used to calculate the Allan deviation of \textbf{a}.
The red trace is behind the blue trace. The locked repetition rate
has much larger frequency excursions than the phase locks of the transfer
lasers. Also visible are the excursions of $f_{225}$ that cause the
deviation from the ideal slope of --1 for the overlapping Allan deviation
shown in \textbf{a}. The counter gate time for this data set is 10
ms.}
\end{figure}

Since the spectral span of our soliton frequency comb is two-thirds
of an octave the \textit{2f-3f} scheme (Fig.1a) can be applied. As
with similar \textit{(n--1)f-nf} schemes \citep{Telle1999,Reichert1999},
the \textit{2f-3f} approach is a trade-off between optical bandwidth
and the requirement of more complex nonlinear conversion. While for
the common \textit{f-2f} scheme a full octave of optical bandwidth
but only one frequency doubling is required, the \textit{2f-3f} scheme
requires only two-thirds of an octave but one frequency doubling and
one frequency tripling. The resulting beat note after the nonlinear conversion
is given by $3(m\cdot f_{\text{{rep}}}+f_{\text{{CEO}}})-2(n\cdot f_{\text{{rep}}}+f_{\text{{CEO}}})=f_{\text{{CEO}}}$
if $2n=3m$ (here $n$ denotes the line number of the doubled frequency
comb line and $m$ the line number of the tripled line) and therefore
enabling the measurement of the carrier envelope offset frequency.
In order to achieve a sufficient signal-to-noise ratio we implement the doubling
and tripling stages using two transfer lasers that are phase locked
to the Kerr frequency comb.

The microresonator used in this work is a silicon nitride waveguide
resonator \citep{Levy2010,Moss2013} with a diameter of $\sim$240
$\mathrm{\mu}$m (Fig.1b), resulting in a free spectral range of $\sim$190
GHz. It is pumped with an amplified external cavity diode laser (ECDL)
operating at $\nu_{\text{{pump}}}\text{\ensuremath{\approx}192.2}$
THz (1560 nm) that is coupled into the chip ($\sim$2 W of cw power
in the waveguide). Using the ``power-kicking'' method \citep{Brasch2016},
the microresonator is brought into a soliton state that gives us directly
from the chip the required bandwidth of two-thirds of an octave (Fig.2a)
\citep{Brasch2016}. The two transfer lasers at $\sim$150 and $\sim$225
THz (2000 and 1330 nm respectively) are phase locked independently
with frequency offsets of $f_{150}$ and $f_{225}$ to their nearest comb
line (Fig.2b). The one transfer laser is then tripled in frequency
(via second harmonic generation followed by sum frequency generation)
while the other is doubled in frequency such that both have a frequency
of around $450$ THz (666 nm) where they generate the desired \textit{2f-3f}
heterodyne beat note ($f_{2f3f}$, Fig.2d).

\begin{figure*}
\includegraphics[width=0.95\textwidth]{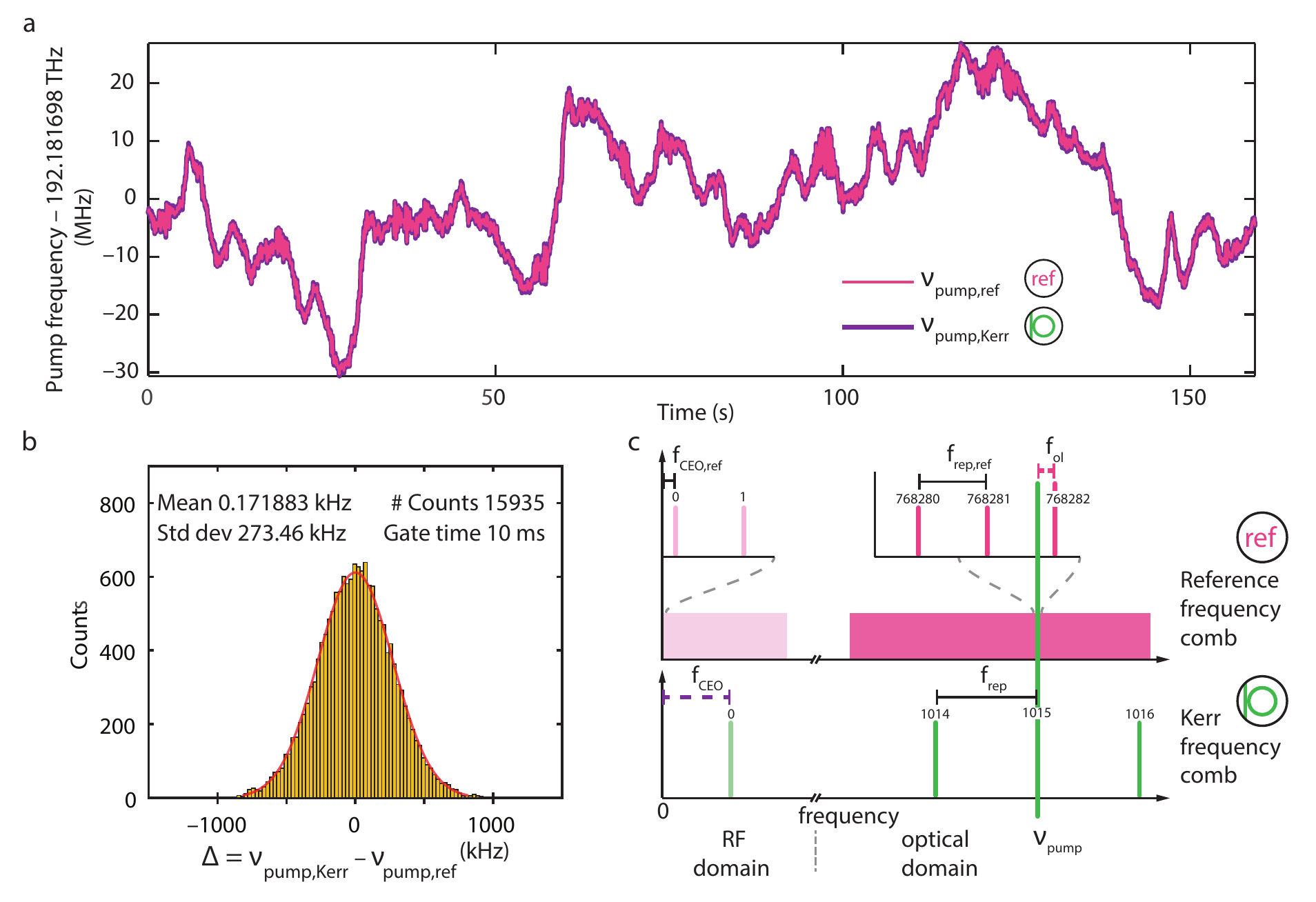}

\caption{\textbf{Tracking of the pump laser frequency with the self-referenced
soliton Kerr frequency comb and an out-of-loop verification. a} The
drift of the frequency of the pump laser ($\nu_{\text{{pump}}}$)
over time as measured with the Kerr frequency comb (thick, dark violet
line, $\nu_{\text{{pump,Kerr}}}=f_{\text{{CEO}}}+1015\cdot f_{\text{{rep}}}$)
and with a reference fiber frequency comb (thin, light pink line,
$\nu_{\text{{pump,ref}}}=f_{\text{{CEO,ref}}}+768282\cdot f_{\text{{rep,ref}}}-f_{\text{{ol}}}$).
Gate time is 10 ms. As the repetition rate of the soliton Kerr frequency
comb is locked and the pump laser itself is one line of the frequency
comb, the drift of the pump laser is equivalent to the drift of the
offset frequency of the Kerr frequency comb. The reference frequency
comb is a fully self-referenced, stabilized fiber frequency comb.
\textbf{b} The histogram of the difference of the two tracked pump
laser frequencies $\Delta=\nu_{\text{{pump,Kerr}}}-\nu_{\text{{pump,ref}}}$.
The Gaussian fit (red) shows a deviation of 172 Hz from the expected mean
of 0 Hz. \textbf{c} Illustration that shows all the frequencies involved
in this out-of-loop measurement. Solid, black horizontal bars indicate
locked frequencies. The two pink and violet dashed bars are the two
frequencies that are not stabilized but counted in order to derive
the data shown in \textbf{a} and \textbf{b}.}
\end{figure*}

Although the large line spacing of our frequency comb of $f_{_{\text{{rep}}}}=189.184$
GHz has the advantage that it can be easily resolved on an optical
spectrometer, one challenge of the large mode spacing is that the
measurement of the repetition rate as well as of the offset frequency
requires the use of high frequency photodiodes and RF components.
In our experiment the repetition rate is measured via optical amplitude
modulation down-mixing \citep{DelHaye2012} and RF down-mixing as
shown in Fig.2c,e. We also take advantage of special properties of
the \textit{2f-3f} scheme with transfer lasers to decrease the measured
frequency $f_{2f3f}$. First, it is important to note that with the
\textit{2f-3f} scheme not all pairs of lines that are doubled and
tripled respectively produce the same $f_{2f3f}$ frequency. There
are two relevant scenarios. The first one is if the condition $2n=3m$
is fulfilled as described above, then the heterodyne beat note $f_{2f3f}$
is equal to $f_{\text{{CEO}}}$. The second scenario is if $2n=3m+1$
is fulfilled, then $f_{2f3f}=f_{\text{{CEO}}}-f_{\text{{rep}}}$ is
measured. Therefore, a pair of lines is chosen to minimize the value
of the measured frequency, which in our case is the second scenario
with $2n=3m+1$. Second, taking into account the two frequency offsets
of the transfer lasers in our setup, the measured $f_{2f3f}$ can
be expressed as $f_{2f3f}=f_{\text{{rep}}}-f_{\text{{CEO}}}+2\cdot f_{\text{{225}}}-3\cdot f_{\text{{150}}}$
\citep{Jost2015}. Therefore we use $f_{150}\approx9.87$ GHz to reduce
the measured \textit{2f-3f} beat to a frequency of the order of 100
MHz (Fig.2d). All frequencies (the repetition rate, the frequency
offsets of the two transfer lasers as well as the \textit{2f-3f} beat
note) are simultaneously monitored on RF frequency counters. These
counters as well as all other RF equipment and in particular the
RF synthesizers for the required local oscillators are referenced
to a common 10 MHz reference derived from a commercial atomic clock
(Fig.2e).

While this is in principle sufficient for self-referencing as the
offset frequency and repetition rate of the frequency comb can be
computed from the counter measurements \citep{Jost2015}, the repetition
rate was also stabilized. For this we compare the repetition rate
of the Kerr frequency comb to the RF reference and feedback onto the
pump power \citep{DelHaye2008,Brasch2016}. We record the overlapping
Allan deviations of the three locked frequencies as shown in Fig.3a. This
is implemented using two gap-less $\Pi$-type counters for the frequencies
$f_{\text{{rep}}}$ and $f_{\text{{225}}}$ as well as one $\Lambda$-type
counter with dead time for $f_{\text{{150}}}$. All overlapping Allan
deviations average down for longer timescales (gate times $\tau>0.1$
s). The flat behavior of the Allan deviation of $f_{\text{{rep}}}$
for shorter timescales is due to the limited bandwidth of the actuation
of its phase lock. However, the slope of --1.10 for longer time scales
matches well the expected value of --1 \citep{Dawkins2007,Bernhardt2009}
showing that the phase lock compensates deviations on these timescales.
In Fig.3b the time traces of the frequency counters are displayed
where the trace of the 225 THz transfer laser shows some smaller excursions.
This results in the slope of --0.54 instead of --1 for this lock in
the overlapping Allan deviation plot. The similar slope of the overlapping
Allan deviation of the 150 THz laser lock however is mainly due to
the different type of counter, which results in a slope close to the
expected --0.5 for this Allan deviation \citep{Dawkins2007,Bernhardt2009}.
Having phase-locked all frequencies but the offset frequency of the
Kerr frequency comb, we can determine the value of the offset frequency
as $f_{\text{{CEO}}}=f_{\text{{rep}}}-3\cdot f_{\text{{150}}}+2\cdot f_{\text{{225}}}-f_{2f3f}\approx159.71$ GHz
and measure its drift by monitoring $f_{2f3f}$.

One unique property of Kerr frequency combs is that the pump laser
constitutes one of the lines of the frequency comb. Because the repetition
rate of our frequency comb is locked, the drift of the pump laser frequency
is directly mapped to the excursion of the offset frequency of the
frequency comb. Therefore the self-referenced Kerr frequency comb
can be used to derive the exact optical frequency of the cw pump laser
and to track it. This is confirmed by taking an out-of-loop measurement.
For this experiment a fraction of the pump laser is split off before
the microresonator and the heterodyne beat note of the pump laser
with a commercial self-referenced, fully stabilized frequency comb
($f_{\text{{CEO,ref}}}=$20 MHz, $f_{\text{{rep,ref}}}=$250.14 MHz)
is counted (Fig.2e). At the same time, the measured $f_{2f3f}$ is
counted as well (Fig.4c). The two frequency counters used are the
same model of $\Pi$-type counters mentioned above and the reference
frequency comb is stabilized to the same commercial atomic clock RF
reference as all other RF equipment used in this experiment. By calculating
the line number of the pump laser in the Kerr frequency comb (1015)
and the line number of the line of the reference frequency comb that
the pump laser beats with (768282) and using our knowledge of all
frequencies ($f_{\text{{rep}}}$, $f_{\text{{CEO}}}$, $f_{\text{{rep,ref}}}$,
$f_{\text{{CEO,ref}}}$ and $f_{\text{{ol}}}$) we can calculate the
optical frequency of the pump laser in two ways. Once using the Kerr
frequency comb and its counted offset frequency and once using the
out-of-loop measurement with the commercial self-referenced fiber
frequency comb. The overlay of these two independent frequency measurements
over time is shown in Fig.4a. The correlation is very clear and no
deviations are visible. In Fig.4b a histogram of the differences between
the two optical frequencies is shown. The data fits well to a Gaussian
distribution with the center frequency of the distribution shifted
by 172 Hz from 0 Hz for the 160-s-long measurement. This out-of-loop
experiment validates our ability to precisely determine the offset
frequency of our Kerr frequency comb using the \textit{2f-3f} scheme.

In summary, we demonstrate a self-referenced Kerr frequency comb without
employing external broadening. Using dissipative Kerr soliton dynamics,
we show that coupling a continuous wave laser into an integrated,
on-chip microresonator is enough to coherently ``broaden'' its spectrum
and to allow for self-referencing. Alleviating the need for additional
external broadening in on-chip Kerr frequency comb devices shows that
self-referenced, phase-stabilized integrated frequency comb sources
are in principle possible. While transfer lasers are used in the current
work, they do in principle constitute elements that are equally amenable
to photonic integration \citep{Chen2016}. Establishing devices that
provide a microwave to optical link on a chip may catalyze a wide
variety of applications such as integrated, microresonator-based atomic
clocks \citep{Papp2014} and on-chip, low-noise RF synthesis from
optical references \citep{Fortier2003} and could contribute to making
frequency metrology ubiquitous.
\begin{acknowledgments}
This work was supported by the European Space Agency (ESA) contract
ESTEC CN 4000108280/12/NL/PA, the Defense Advanced Research Projects
Agency (DARPA) contract W911NF-11-1-0202 (QuASAR) and the Swiss National
Science Foundation. This material is based on work supported by the
Air Force Office of Scientific Research, Air Force Material Command,
under award FA9550-15-1-0099. V.B. acknowledges support from the ESA
via contract ESTEC CN 4000105962/12/NL/PA. J.J. acknowledges support
by the Marie Curie IIF Fellowship. M.G. acknowledges support from
the Hasler foundation and support from the ‘EPFL Fellows’ fellowship
program co-funded by Marie Curie, FP7 Grant agreement no. 291771.
The sample was fabricated at the Centre for MicroNanotechnology (CMi)
at EPFL. 
\end{acknowledgments}

\bibliographystyle{apsrev4-1}
\bibliography{2f3fArxiv}

\end{document}